\documentstyle[epsfig,aps,twocolumn]{revtex}

\newcommand{\unit}{\hat{\bf n}}

\newcommand{\rv}{{\bf r}}
\newcommand{\Ev}{{\bf E}}

\newcommand{\kv}{{\bf k}}

\newcommand{\beq}{\begin{equation}}
\newcommand{\eeq}{\end{equation}}
\newcommand{\bea}{\begin{eqnarray}}
\newcommand{\eea}{\end{eqnarray}}

\newcommand{\<}{\langle}
\renewcommand{\>}{\rangle}

\newcommand{\commentout}[1]{{}}

\begin{document}

\draft
\preprint{}
\title{Schr\"odinger cat state of a Bose-Einstein
condensate in a double-well potential}
\author{J. Ruostekoski}
\address{Abteilung f\"ur Quantenphysik,
Universit\"at Ulm, D-89069 Ulm, Germany}
\date{\today}
\maketitle
\begin{abstract}
We consider a weakly interacting coherently coupled Bose-Einstein condensate
in a double-well potential. We show by means of stochastic simulations
that the system could possibly be driven to an entangled macroscopic 
superposition
state or a Schr\"odinger cat state by means of a continuous quantum
measurement process.
\end{abstract}

\section{Introduction}

Since the first observations of Bose-Einstein condensation in dilute
alkali-metal atomic gases \cite{AND95,BRA95,DAV95} the ultra-cold
atomic gases have stimulated significant theoretical and experimental
interest \cite{PAR98}. The scientific progress has been rapid and
examples of recent experiments include the development of accurate detection
methods \cite{STA99}, the state preparation of topological structures
\cite{MAT99}, and the applications in nonlinear atom optics \cite{DEN99}. 
Due to the macroscopic quantum
coherence Bose-Einstein condensates (BECs) could possibly be also
used in the future as a test for the foundations of quantum mechanics.
One particularly puzzling and controversial issue has been the existence
of macroscopic quantum superposition states in many-particle quantum systems.
In this paper we propose a method of creating the Schr\"odinger cat states
of different atom occupation numbers in a weakly interacting BEC
confined in a double-well potential.

The existence of the superpositions of macroscopically distinguishable 
states in BECs has been addressed by several authors 
\cite{RUO98a,CIR98,STE98,GOR99,DAL00}. The superposition
state may arise as the ground state of a coherently coupled BEC in 
a double-well potential \cite{CIR98,STE98}. 
Under certain conditions it could be reached
as a result of a unitary time evolution \cite{GOR99}. 
Previously, we proposed a method
of creating Schr\"odinger cat states in BECs by means of scattering light
from two BECs moving with opposite velocities \cite{RUO98a}. 
The nonunitary evolution
due to the detections of scattered photons drives the condensates to
macroscopic quantum superposition states. In this paper we show that
a continuous quantum measurement process could also drive a trapped
coherently coupled BEC in a double-well potential to a Schr\"odinger 
cat state.
The advantage of the proposed scheme is that the BEC is almost
stationary and trapped. Moreover, as a result of the back-action
of quantum measurement process the superposition state 
could be reached rapidly unlike in a slow unitary evolution, which
may be very sensitive to decoherence.

The paper is organized as follows: We begin in Sec.~\ref{uni}
by introducing the unitary system Hamiltonian. The scattering
of light and the measurement geometry is described in Sec.~\ref{light}.
In Sec.~\ref{stoc} we study the dynamics of the open quantum system 
in terms of stochastic trajectories of state vectors. The results
of the numerical simulations are presented in Sec.~\ref{num}.
Finally, a few concluding remarks are made in Sec.~\ref{rem}.

\section{System dynamics}

\subsection{Unitary evolution}
\label{uni}

We consider the evolution of a BEC in a double-well potential in 
a two-mode approximation. Macroscopic quantum coherence of BECs 
results in coherent quantum tunneling of atoms between the two modes 
representing `two BECs'. This is analogous to the coherent tunneling 
of Cooper pairs in a Josephson junction 
\cite{JAV86,MIL97,SME97,ZAP98,RUO98b,WIL99}.
To obtain the system Hamiltonian in the two-mode approximation
for the unitary evolution of the BEC we approximate the total field 
operator by the
two lowest quantum modes $\psi(\rv)\simeq \psi_b(\rv) b +\psi_c(\rv)c$, where
$\psi_b$ and $\psi_c$ stand for the local mode solutions of the individual 
wells with small spatial overlap. The corresponding annihilation operators 
are denoted by $b$ and $c$. The Hamiltonian in the two-mode approximation 
reads \cite{MIL97}:
\beq
{H_S\over\hbar}=\xi b^\dagger b+\Omega (b^\dagger c + c^\dagger b)+
\kappa [ (b^\dagger)^2 b^2 + (c^\dagger)^2 c^2]\,.
\label{eq:ham}
\eeq
Here $\xi$ is the energy difference between the modes. The tunneling
between the two wells is described by $\Omega$, which is proportional to the
overlap of the spatial mode function of the opposite wells. The short-ranged
two-body interaction strength is obtained from $\kappa=2\pi a\hbar /m \int
|\psi_b (\rv)|^4$, where $a$ and $m$ denote the scattering length and
the atomic mass, respectively. For simplicity, here we have assumed that
$\int |\psi_b (\rv)|^4= \int |\psi_c (\rv)|^4$.
A necessary condition for the
validity of the single-mode approximation in a harmonic trap is that
the oscillation energy of the atoms does not dominate over the mode energy
spacing of the trap.

According to the Josephson effect, the 
atom numbers of the BECs determined by the Hamiltonian {(\ref{eq:ham})}
may oscillate even if the number of atoms in each well is initially equal. 
Due to the nonlinear self-interaction
the number oscillations also exhibit collapses and revivals.
These have been studied numerically in Ref.~\cite{MIL97}. We may
also obtain a simple analytical description by solving the dynamics
in the rotating wave approximation in the limit $\Omega\gg N\kappa$
as described in Ref.~\cite{KOR97}. Here $N$ denotes the total number 
of atoms. In particular, we may solve the number of atoms $N_b\equiv\< 
b^\dagger b\>$ in well $b$. We consider a coherent state in the both
wells as an initial state. Then we obtain
\beq
N_b = {N\over2}\big[1+e^{N[\cos(\kappa t)-1]}
(\sqrt{1-\beta^2}\cos\eta
-\beta\sin\varphi\sin\eta)\big]\,,
\label{atomnum}
\eeq
with $\eta\equiv N\beta \sin(\kappa t)\cos\varphi-2\Omega t$.
Here all the operators on the right-hand side have been evaluated at $t=0$.
It is useful to define the real expectation values $\beta$ and $\varphi$
in the following way:
\beq
\beta e^{i\varphi}\equiv {2\over N} \< b^\dagger c \>\,.
\label{bet}
\eeq
For a coherent state with equal atom numbers in the two wells
we obtain the visibility $\beta=1$. The relative phase between the wells is
$\varphi$. For a number state there is no phase information and
$\beta=0$. For unequal atom numbers the maximum visibility is
$\beta_{\rm max}=2(N_bN_c)^{1/2}/N$. We see that the number of atoms
in Eq.~{(\ref{atomnum})} may oscillate in the case of initially
equal atom numbers $\beta=1$. The amplitude of sinusoidal oscillations,
representing the macroscopic coherence, collapses. For instance, for $\varphi=
\pi/2$ and $\beta=1$ we may obtain the short time decay by considering
the time scales $N\kappa t\ll 1\ll \Omega t$. Then the decay of the
oscillations has the form $\exp(-N\kappa^2 t^2/2)$.
This is the rate of the phase diffusion and it may be interpreted
as the width of the relative phase $\langle\Delta\varphi^2(t)
\rangle\simeq N\kappa^2 t^2\sim \kappa^2
t^2/  \langle\Delta\varphi^2(0)\rangle$. Perhaps surprisingly the
functional dependence of the width in this case is the
same as in the case of two uncoupled BECs~\cite{PAR98}.

If the phase is unknown we may obtain the ensemble average by
integrating over the relative phase $\varphi$ in Eq.~{(\ref{atomnum})}:
\bea
N_b &=&{N\over2}\big\{ 1+e^{N[\cos(\kappa t)-1]}\sqrt{1-\beta^2}
\nonumber\\
&&\times\cos(2\Omega t) \, J_0[N\beta\sin(\kappa t)]\big\}\,, 
\label{eq:numc3}
\eea
where $J_0$ is the 0th order Bessel function. If the both wells have
initially equal number of atoms, $\beta=1$, the atom numbers do not
oscillate. For unequal atom numbers the oscillations collapse at the
first zero of the Bessel function $t\simeq 2.4/(N\beta\kappa)$ for
$(N_b N_c)^{1/2}\gg 1$.

\subsection{Quantum measurement process}
\label{light}

\begin{figure}
\begin{center}
\leavevmode
\epsfig{
width=6cm,file=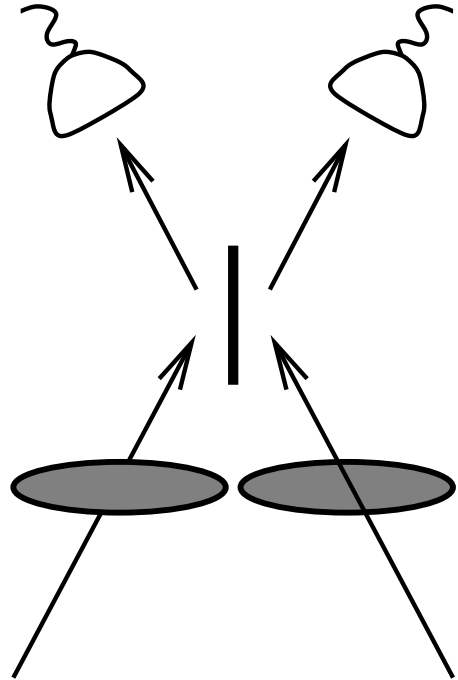}
\end{center}
\caption{
The measurement setup. Two incoming light fields are scattered
from two coherently coupled BECs. The two scattered photon beams are
combined by a 50-50 beam splitter. The photons are detected from
the two output channels of the beam splitter. One of the output
channels introduces only a constant phase shift and it may be ignored.
}
\label{fig1}
\end{figure}

The time evolution of the system is nonunitary, when we include the
effect of quantum measurement process. We consider the nondestructive 
measurement of the number of atoms in the both wells by means of
shining coherent light beams
through the atom clouds. The scattered light beams are combined by
a 50-50 beam splitter. We display the measurement setup in
Fig.~\ref{fig1}.

We assume that the incoming light fields are detuned far from the atomic
resonance. For instance, if the shape of the gas is
flat and the light is shone through a thin dimension, the multiple scattering
is negligible and the sample can be considered optically thin.
A BEC atom scatters back to the BEC via coherent spontaneous scattering, 
stimulated by a large number of atoms in the BEC. 
Coherently scattered photons are emitted into a
narrow cone in the forward direction. By spontaneous
scattering we mean that the emission is not stimulated by light, 
although it is stimulated by atoms. The decay into noncondensate 
center-of-mass states is also stimulated by the Bose-Einstein 
statistics. However, at very low temperatures this stimulation
is much weaker because most of the particles are in the BEC.
As a first approximation we ignore the scattering from and to the
noncondensate modes.
Then the measurement is nondestructive 
in the sense that BEC atoms in the modes $b$ and $c$ scatter back to 
the same modes $b$ and $c$. Because the overlap of the mode functions
of the different wells is assumed to be small, the scattering between
the two wells is ignored.

The detection rate of photons on the detectors is
the intensity of the scattered light $I(\rv)$
integrated over the scattering directions divided 
by the energy of a photon $\hbar c k$. 
Here $k$ and $c$ stand for the wave number and the velocity of 
light. We obtain the 
detection rate at the channel $j$:
\beq
\gamma_j = {1\over \hbar c k}\int d\Omega_{\unit}\, r^2
I_j(\rv) =2 \Gamma\langle C_j^\dagger C_j \rangle\,.
\label{rate}
\eeq
The photon annihilation operator at the output channel $j$ of the
beam splitter is denoted by $C_j$. For a symmetric measurement
geometry we obtain
\beq
C_1 =  {1\over\sqrt{2}} (b^\dagger b-c^\dagger c),\quad
C_2 =  {1\over\sqrt{2}} (b^\dagger b+c^\dagger c)\,.
\label{eq:relaxch}
\eeq
Because the total number of atoms is assumed to be conserved,
the operator $\hat{N}\equiv b^\dagger b+c^\dagger c$ contributes to
the measurements only through a constant phase shift. Therefore,
we may ignore the effect of the scattering channel 2 on the dynamics.

The scattered intensity may be written in terms of the
positive frequency component of the scattered electric field $\Ev^+(\rv)$
\beq
I(\rv )=2c\epsilon_0\langle\Ev^-(\rv)
\cdot\Ev^+(\rv)\rangle\,,
\eeq
Here $\epsilon_0$ denotes the permittivity of the vacuum.

We assume that the driving electric fields may be approximated by
plane waves ${\bf E}^+_{d}({\bf r}) ={\cal E} \hat{\bf e}\,
e^{i(\kv\cdot\rv-\Omega t)}/2$. In the limit of large atom-light
detuning $\Delta$ we use the first Born approximation
and write the electric fields in the far radiation 
zone ($kr\gg1$). Then the scattered field from the well $b$ has 
the following form \cite{JAV95b}:
\bea
\Ev^+(\unit r) &=& {k^2{\cal R}e^{ikr}\over 4\pi \epsilon_0\Delta r}\,
\unit\times(\unit\times{\bf d})\nonumber\\
&& \int d^3r' e^{i(\kv-k\unit)\cdot\rv'}
|\psi_b(\rv')|^2 b^\dagger b\,.
\label{sfield}
\eea
Here we have defined the Rabi frequency ${\cal R}$ of the atomic dipole
matrix element ${\bf d}$ by $ {\cal R}\equiv {d {\cal E}/ (2\hbar)}$.
We also assumed that ${\bf d}\cdot\hat{\bf e}=d$.
In the limit that the characteristic
length scale $\ell$ of the BECs is much larger than the inverse of the
wave number of the incoming light $\ell\gg1/k$, the momentum of the
scattered photon is approximately conserved, and we obtain in
Eq.~{(\ref{sfield})}:
$$
\int d^3r' e^{i(\kv-k\unit)\cdot\rv'}
|\psi_b(\rv')|^2 \simeq  \delta(k\unit-\kv)
$$
In this simple case the scattering rate $\Gamma$ may be easily
evaluated:
\beq
\Gamma = {3\gamma{\cal R}^2\over 8\pi\Delta^2}\,,
\eeq
Here $\gamma={d^2 k^3/ (6\pi\hbar\epsilon_0)}$ denotes the optical 
linewidth of the atom.

\subsection{Stochastic Schr\"odinger equation}
\label{stoc}

The dissipation of energy from the quantum system of macroscopic light 
fields and the BEC in a double-well potential is described by the 
coupling to a zero 
temperature reservoir of vacuum modes, resulting in a spontaneous 
emission linewidth for the atoms.
The dynamics of the continuous quantum measurement process may be
unraveled into stochastic trajectories of state vectors 
\cite{DAL92,GAR92,CAR93}. The procedure consists of the evolution 
of the system with a non-Hermitian Hamiltonian $H_{\rm eff}$, 
and randomly decided quantum `jumps'.
In our case the quantum jumps correspond to the
detections of spontaneously emitted photons.
The system evolution is thus conditioned on the outcome 
of a measurement. The non-Hermitian Hamiltonian has the following
form:
\beq
H_{\rm eff}=H_S-i\hbar \Gamma C^\dagger_1C_1\,,
\label{eq:nonher}
\eeq
where the unitary system Hamiltonian $H_S$ is determined 
by Eq.~{(\ref{eq:ham})}.

Equation {(\ref{eq:nonher})} corresponds to the modification of the 
state of the system associated with a zero detection result 
for scattered photons. Because the output is being continuously monitored, 
we gain information about the system even if no photons have been detected.

The Hamiltonian $H_{\rm eff}$ determines the evolution of the state
vector $\psi_{\rm sys}(t)$. If the wave function $\psi_{\rm sys}(t)$ is
normalized, the probability that a photon from the output channel 1
of the beamsplitter is detected during the time interval $[t,t+\delta t]$
is
\beq
P(t)=2\Gamma\langle\psi_{\rm sys}(t)\, |\, C^\dagger_1C_1\, |\, \psi_{\rm
sys}(t)\rangle\, \delta t\,.
\label{eq:prob1}
\eeq

We implement the simulation algorithm as follows:
At the time $t_0$ we generate a quasi-random number
$\epsilon$ which is uniformly distributed between 0 and 1. We assume that the
state vector $\psi_{\rm sys}(t_0)$ at the time $t_0$ is normalized. Then we
evolve the state vector by the non-Hermitian Hamiltonian $H_{\rm eff}$
iteratively for finite time steps $\Delta t$. At each time step $n$ 
we compare
$\epsilon$ to the reduced norm of the wave function, until 
$$\langle \psi_{\rm sys}(t_0+n\Delta t)\, |\, 
\psi_{\rm sys}(t_0+n\Delta t )\rangle <\epsilon\,,$$ 
when the detection of a photon occurs. If the photon has been
observed during the time step $t\rightarrow t+\Delta t$ we take the new wave
function at $t+\Delta t$ to be
\beq
|\, \psi_{\rm
sys}(t+\Delta t)\rangle=\sqrt{2\Gamma}\,C_1 \,|\, \psi_{\rm
sys}(t)\rangle\,,
\label{eq:meas1}
\eeq
which is then normalized.

\section{Numerical results}
\label{num}

We simulate the effect of the system Hamiltonian $H_S$ and the quantum
measurement process of scattered photons by means of the stochastic
Schr\"odinger equation. For simplicity, we set the total number of 
atoms to be reasonably small $N=200$.
We start from a slightly asymmetric initial state with the two modes
in number states $N_b=102$ and $N_c=98$. We choose $\Gamma/\Omega=
5\times10^{-6}$.

\begin{figure}
\begin{center}
\leavevmode
\begin{minipage}{4.2cm}
\epsfig{
width=4.2cm,file=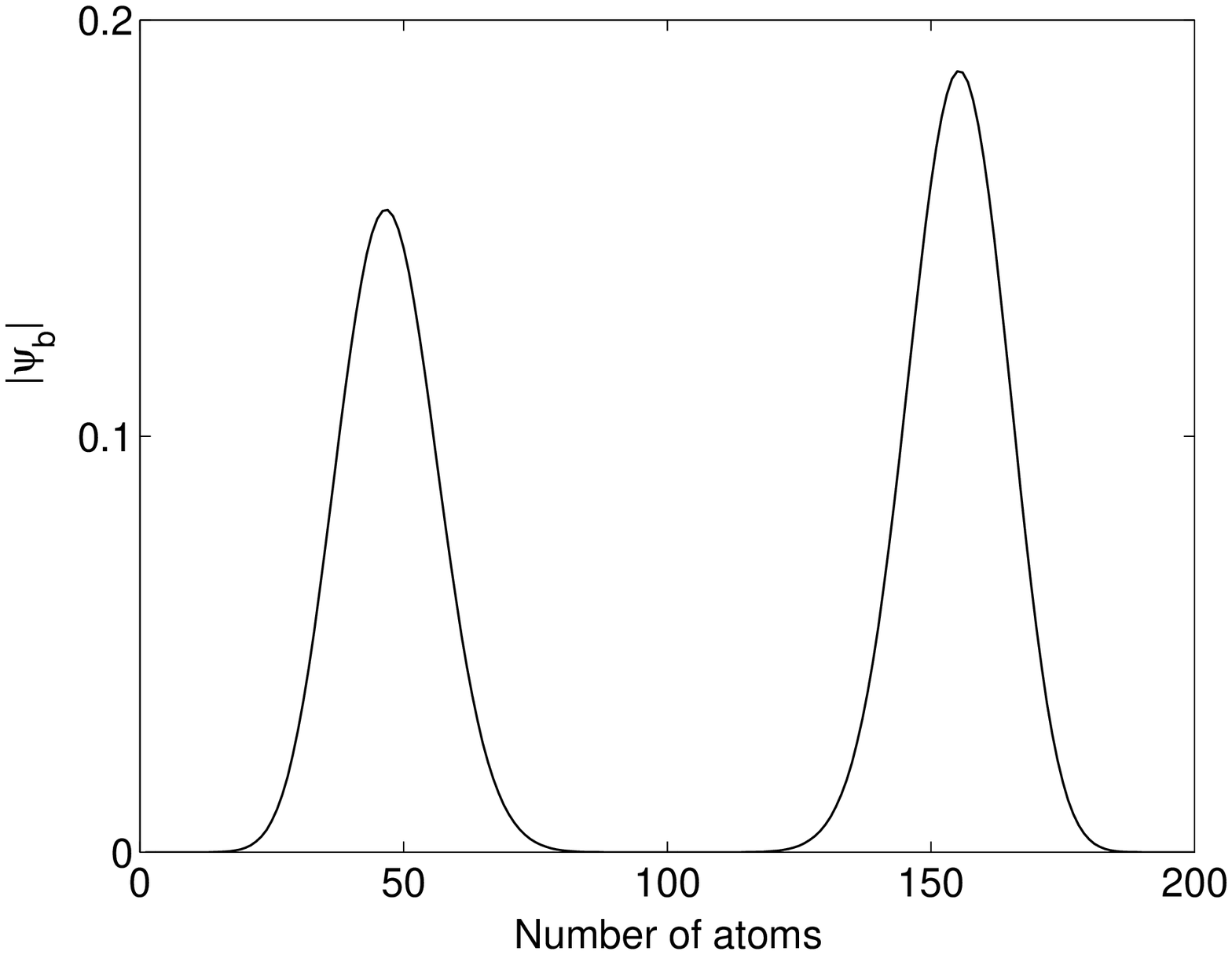}
\end{minipage}
\begin{minipage}{4.2cm}
\epsfig{
width=4.2cm,file=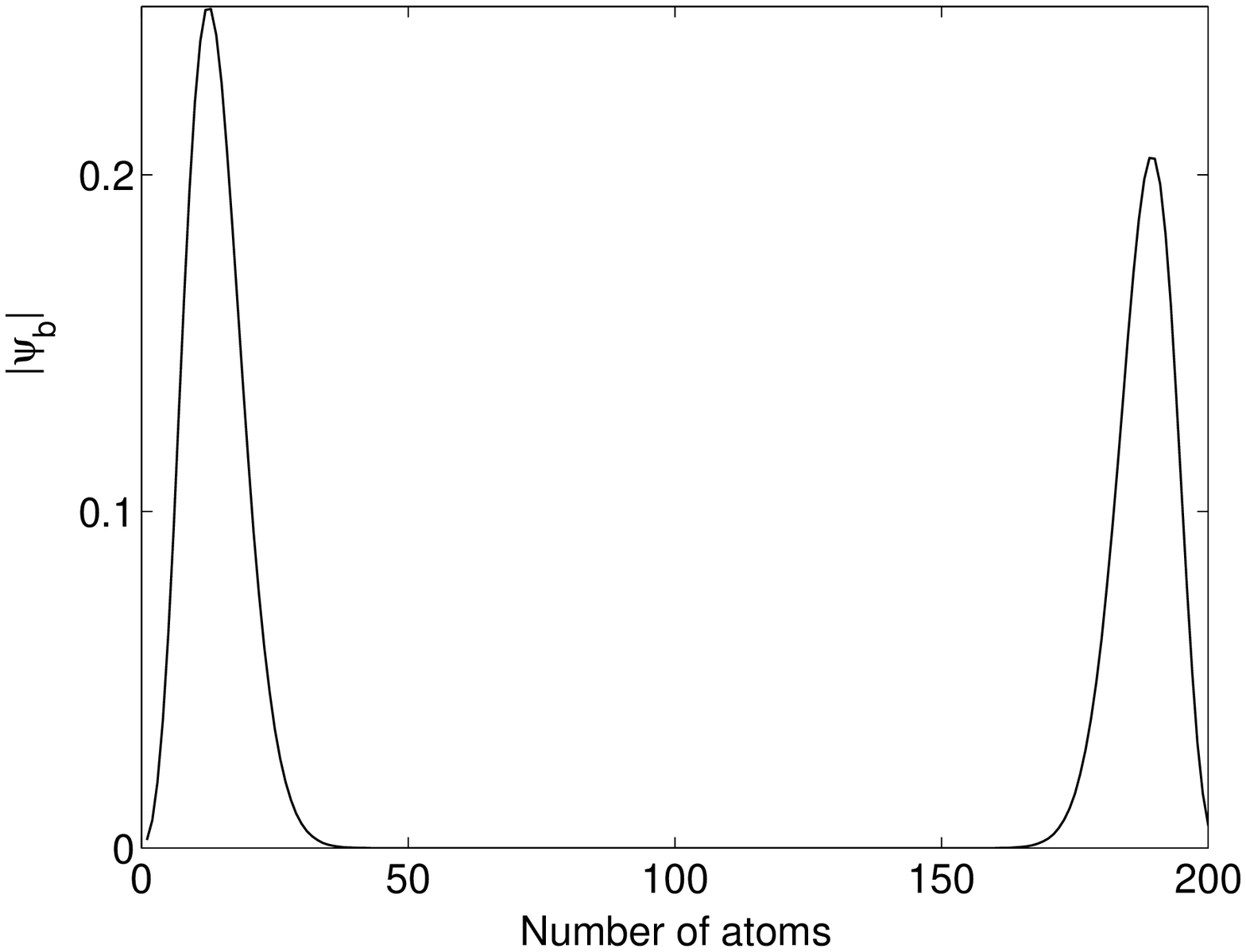}
\end{minipage}
\end{center}
\caption{
The entangled Schr\"odinger cat states of atoms with different atom numbers.  
We show the absolute value of the wave function $|\psi_b|$ in the number 
state basis for the atoms in well $b$ during one realization of stochastic
measurement process of spontaneously scattered photons after (a) 750
and after (b) 1700 detections.
The two maxima correspond to the superposition states. 
The nonlinearity $\kappa=0$ and the total number of atoms $N=200$. 
}
\label{fig2}
\end{figure}

\begin{figure}
\begin{center}
\leavevmode
\begin{minipage}{4.2cm}
\epsfig{
width=4.2cm,file=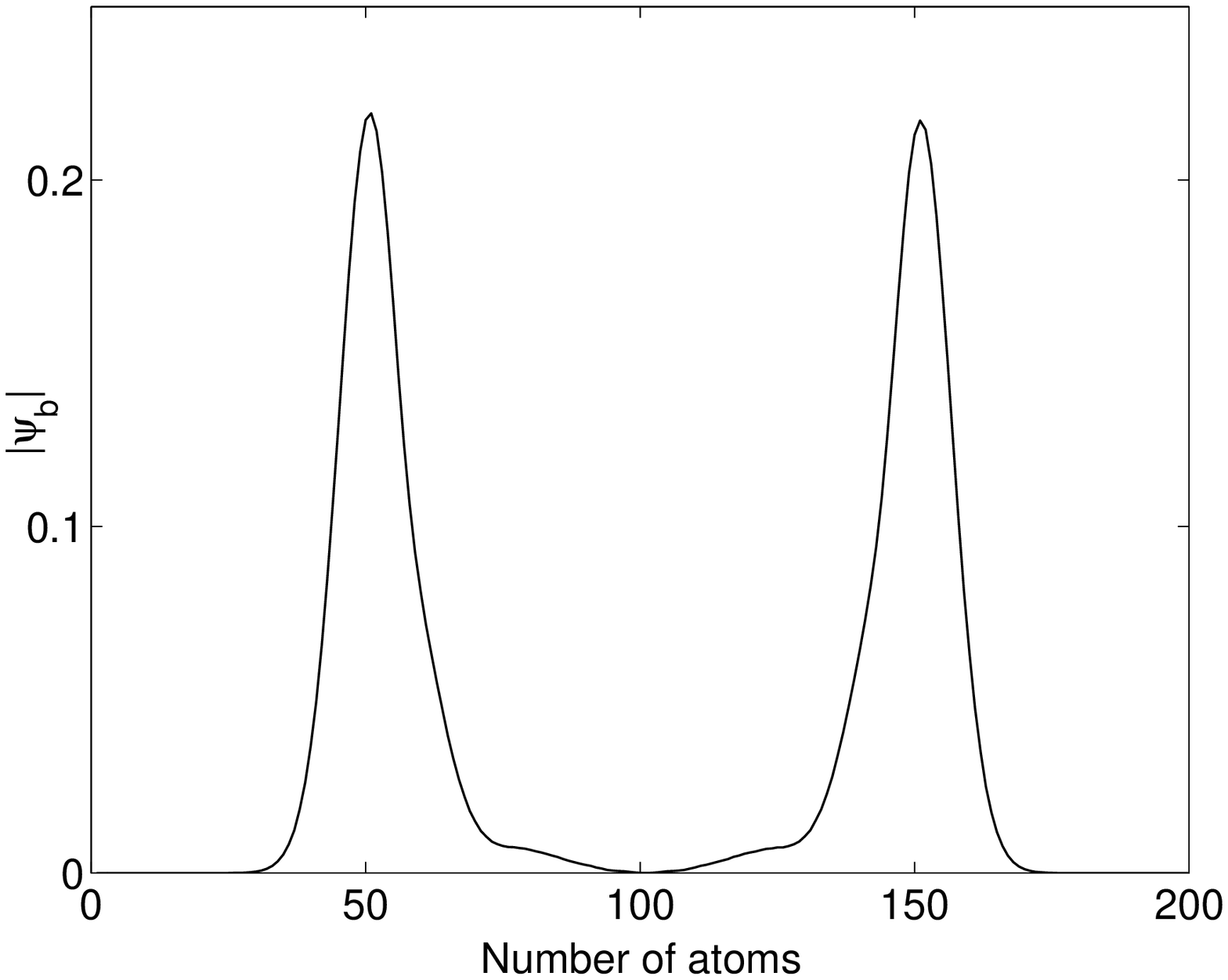}
\end{minipage}
\begin{minipage}{4.2cm}
\epsfig{
width=4.2cm,file=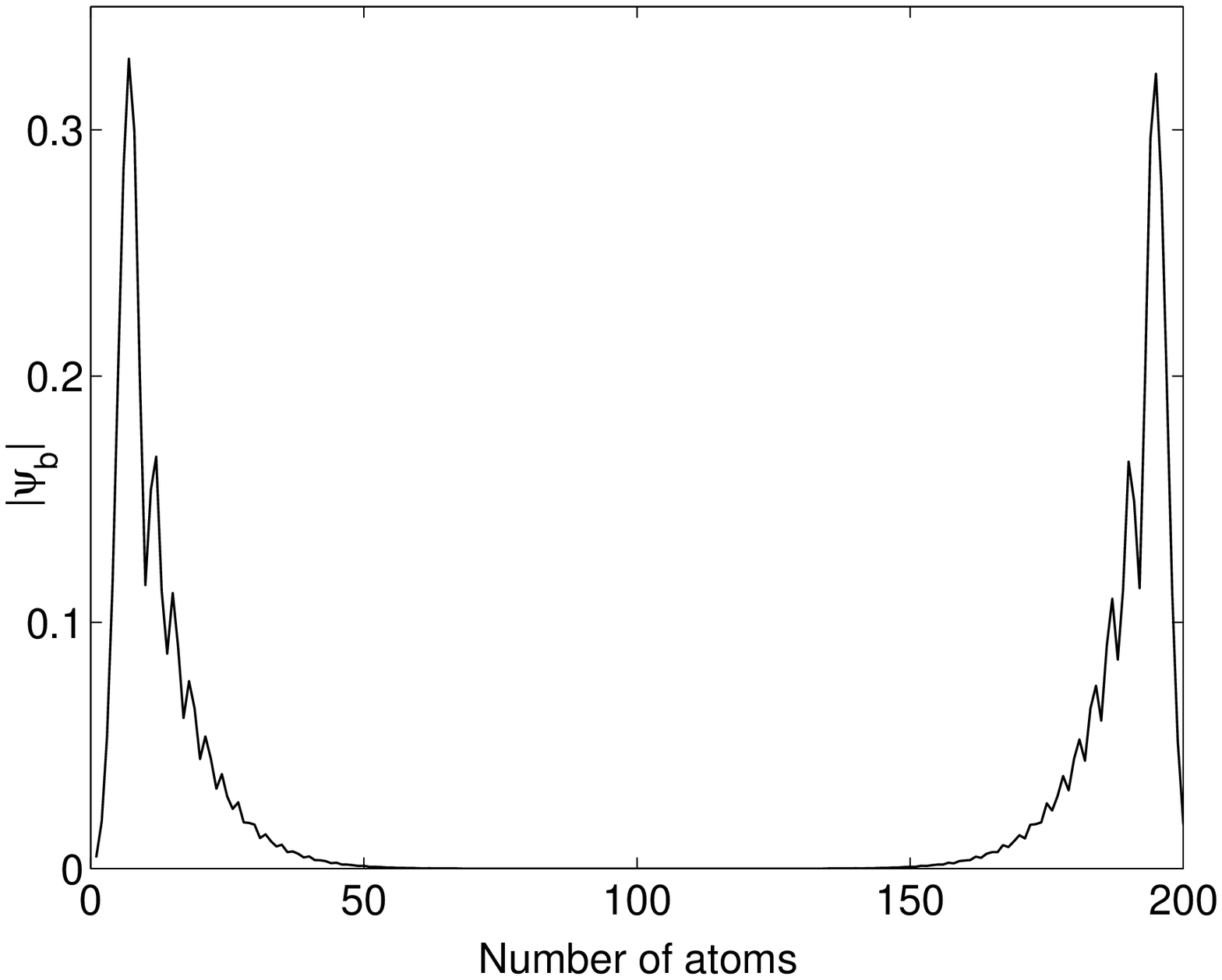}
\end{minipage}
\end{center}
\caption{
The entangled Schr\"odinger cat states of atoms with different atom numbers.  
We show the absolute value of the wave function $|\psi_b|$ 
during one realization of stochastic trajectory after (a) 100
and after (b) 1250 detections. The nonlinearity $N\kappa=0.2$.
}
\label{fig2b}
\end{figure}

After just a few detected photons we observe the
emergence of two well-separated amplitude maxima in the occupation
number of atoms in one of the two wells. These correspond to a macroscopic
number state superposition or a Schr\"odinger cat state. Because the
total number of atoms is assumed to be conserved, the atom numbers in the
two wells are entangled and we have a Bell-type of superposition state.
In Fig.~\ref{fig2} we display the absolute value of the wave function 
$|\psi_b|$ in mode $b$ in number state basis in a single run at two 
different times. In this case the nonlinearity
vanishes $\kappa=0$ and $\xi/\Omega=0.1$. 
We clearly recognize the two distinct 
peaks in the number distribution. For instance, the peaks in 
Fig.~\ref{fig2} (b) are centered at
$N_b\simeq 10$ and $N_b\simeq 190$ corresponding
to maxima at $N_c\simeq 190$ and $N_c\simeq 10$,
respectively. In Fig.~\ref{fig2b} we show the absolute value of the
wave function for a different run with $N\kappa/\Omega=0.2$ and 
$\xi/\Omega=0.001$.

\begin{figure}
\begin{center}
\leavevmode
\epsfig{
width=6cm,file=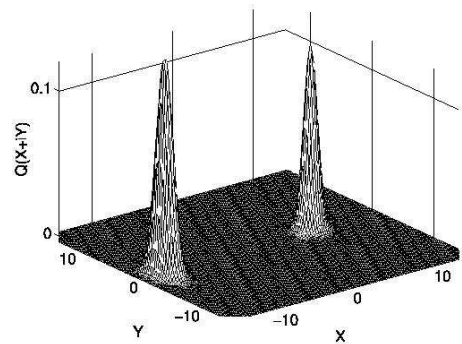}
\end{center}
\caption{
The $Q$ function of the Schr\"odinger cat state.
We show the $Q$ function of the quantum state displayed in Fig.~\ref{fig2}
(a). The two peaks are located at different value of the radius
$N_b^{1/2}$ representing the different maxima of the occupation numbers.
}
\label{fig3}
\end{figure}

We also describe the state of the BEC in terms of the quasiprobability $Q$
distribution.
For the number state distribution of atoms $|\psi_b\rangle = \sum_n c_n
|n\rangle$ in mode $b$ we obtain \cite{WAL94}:
\beq
Q(\alpha)={ |\langle\alpha |\psi_b \rangle |^2 \over\pi} =
{e^{-|\alpha|^2}\over\pi}\left|\sum_{n=0}^N {\alpha^n c^*_n\over
\sqrt{n!}}\right|^2\,.
\label{eq:Q}
\eeq
The $Q$ function represents the phase-space distribution. The
amplitude and phase quadratures are denoted by $X$ and $Y$. In polar
coordinates the radius in the $xy$ plane is equal to $N_b^{1/2}$ and the polar
angle is the relative phase between the atoms in the two wells.
In Fig.~\ref{fig3} we show the $Q$ function distribution of the number
state superposition displayed in Fig.~\ref{fig2}(a).

\begin{figure}
\begin{center}
\leavevmode
\begin{minipage}{4.2cm}
\epsfig{
width=4.2cm,file=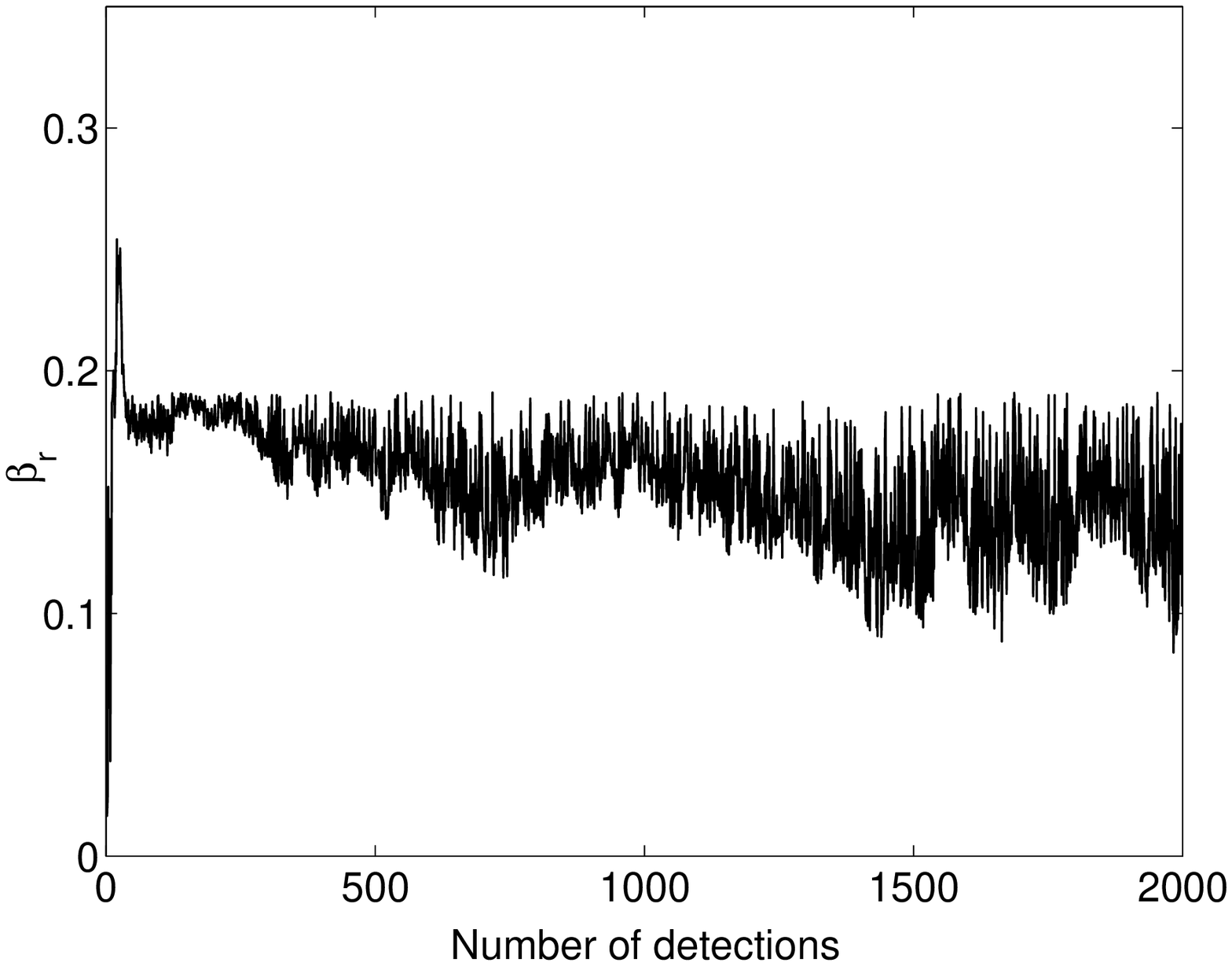}
\end{minipage}
\begin{minipage}{4.2cm}
\epsfig{
width=4.2cm,file=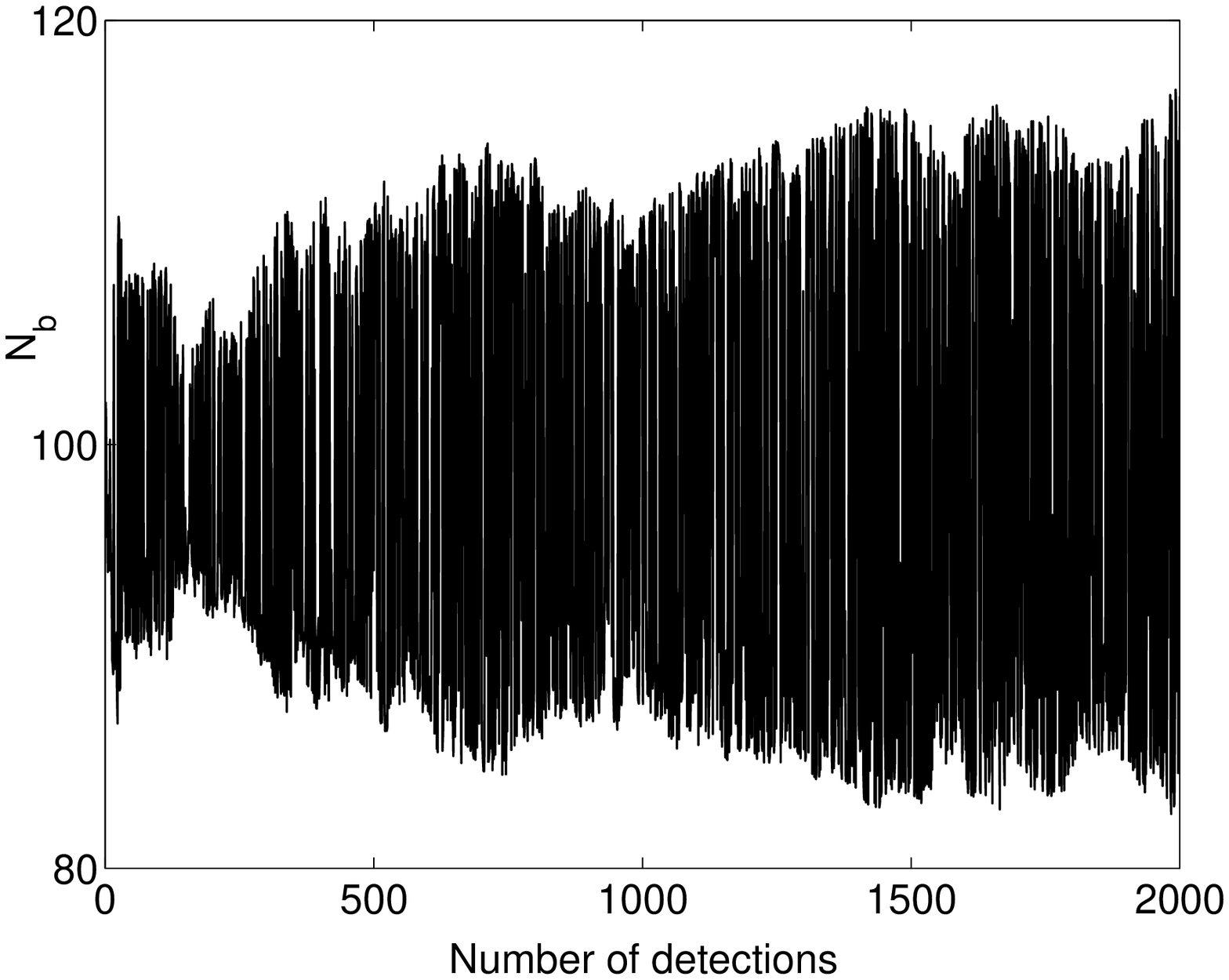}
\end{minipage}
\end{center}
\caption{
The (a) relative visibility of the interference $\beta_r$ and the 
(b) number of atoms $N_b$ in one of the wells as a function of the
number of detected photons. Due to the Schr\"odinger cat state the
visibility never reaches one. Here the nonlinearity $\kappa=0$.
}
\label{fig4}
\end{figure}

\begin{figure}
\begin{center}
\leavevmode
\begin{minipage}{4.2cm}
\epsfig{
width=4.2cm,file=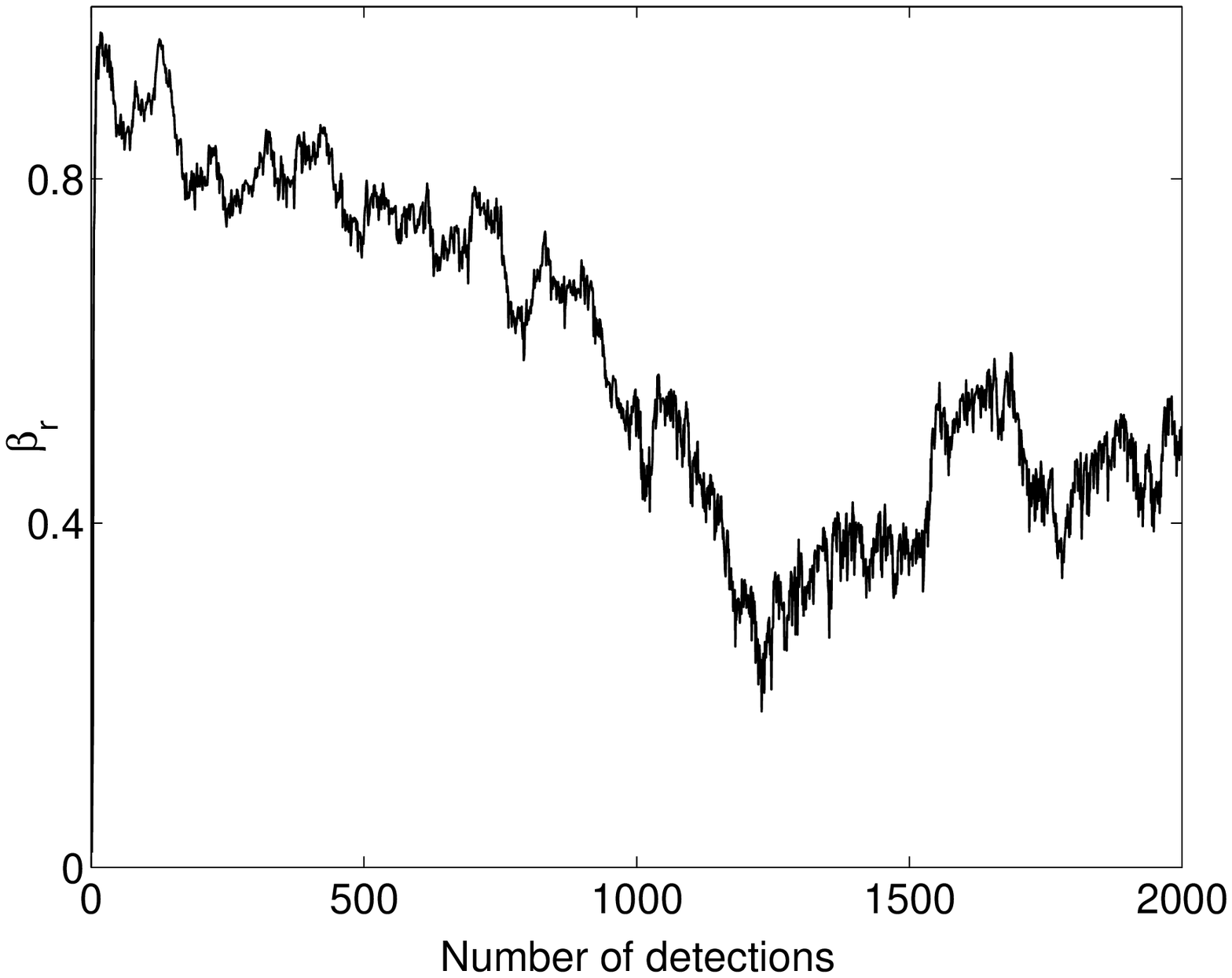}
\end{minipage}
\begin{minipage}{4.2cm}
\epsfig{
width=4.2cm,file=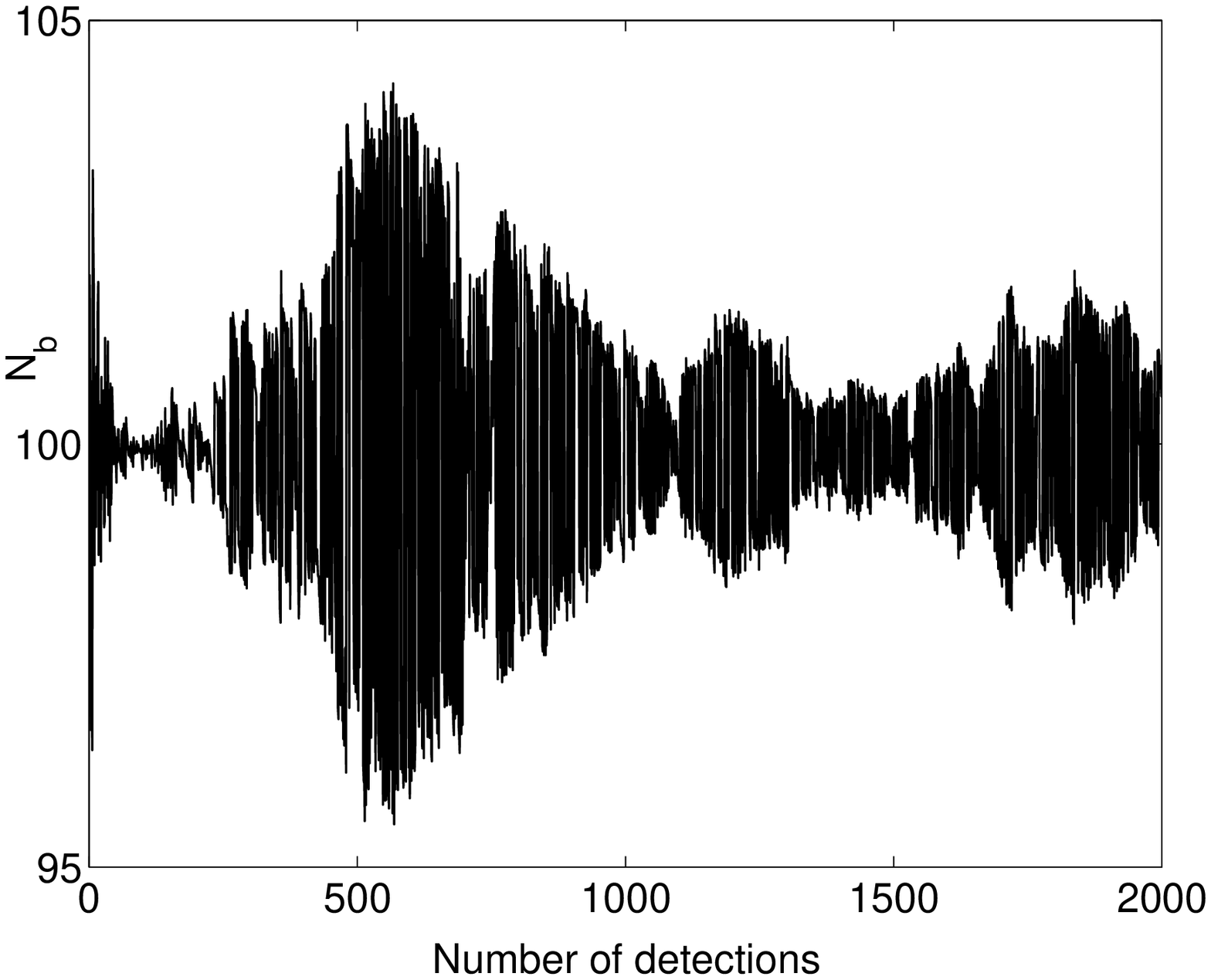}
\end{minipage}
\end{center}
\caption{
The (a) relative visibility of the interference $\beta_r$ and the 
(b) number of atoms $N_b$ in one of the wells as a function of the
number of detected photons. The nonlinearity $N\kappa/\Omega=0.2$.
}
\label{fig5}
\end{figure}

It is interesting to emphasize that the measurement of the number of atoms
in only one of the wells affects the system dynamics quite differently.
In that case the number state distribution remains well localized and
approximately approaches a coherent state \cite{RUO98b}. 
Even though we start from
a number state with no phase information, the detections of spontaneously
scattered photons establish a macroscopic coherence or the off-diagonal
long-range order (ODLRO) between the atoms in the two separate wells.
This is similar to establishing the coherence between two BECs as
a result of the counting of atoms \cite{JAV96,JAC96}.
However, in the present case the continuous measurement process drives
the system to a Schr\"odinger cat state and the ODLRO remains small.
We may describe the visibility of the macroscopic coherence between 
the two wells by the real parameter $\beta$ defined in Eq.~(\ref{atomnum}).
We show the relative visibility
$\beta_r\equiv \beta/\beta_{\rm max}$ and the number of atoms in
well $b$ as a function of the number of measurements for $\kappa=0$ in
Fig.~\ref{fig4} and for $N\kappa/\Omega=0.2$ in Fig.~\ref{fig5}.
Due to the emergence of the superposition state
the visibility remains below one. The measurement process of
the scattered photons significantly complicates the dynamics of the
number of atoms predicted by the unitary time evolution of
Eq.~{(\ref{eq:ham})}.

\section{Final remarks}
\label{rem}

We studied the generation of the macroscopic superposition states or
the Schr\"odinger cat states of a BEC in a double-well potential. The
Schr\"odinger cat state was shown to emerge as a result of the
continuous quantum measurement process of scattered photons. 
The particular detection geometry increses the fluctuations
of the relative atom number between the two wells. Therefore the
superposition states are more stable in the detection
process. The proposed setup is an open quantum system and the creation
of the Schr\"odinger cat state in this case is not based on reaching 
the ground state of a BEC in a double-well potential \cite{CIR98,STE98}.
The advantage over previously proposed open systems schemes \cite{RUO98a}
is that the BEC is stably trapped and the superposition state for a
small BEC could be created by scattering only a few photons.

In the present discussion we ignored the effect of decoherence
\cite{ZUR91}. The interaction of the BEC with its environment
results in the decoherence of the superposition states. We can identify
several sources of decoherence. Decoherence
by amplitude damping or by phase damping has been estimated in
Ref.~\cite{WAL85}. The inelastic two-body and three-body collisions 
between the 
condensate atoms and the noncondensate atoms change the number of 
condensate atoms and introduce amplitude damping. The phase damping 
corresponds, e.g., to elastic collisions between the condensate and 
noncondensate atoms in which case the number of BEC atoms is conserved.
If the number of atoms in a BEC is not large, the scattering between
the condensate and noncondensate atom fractions may not be negligible.
This also introduces amplitude decoherence. Additional sources of
decoherence may be, e.g., the imperfect detection of the scattered
photons and the fluctuations of the magnetic trap.
In Ref.~\cite{DAL00} it was proposed that the decoherence rate of
a BEC could be dramatically reduced by symmetrization of the
environment and by changing the geometry of the trapping potential
to reduce the size of the thermal cloud. Moreover, the continuous 
measurement process increases the information about the system and 
therefore it could also reduce the decoherence rate.

\section*{Acknowledgements}

We are indebted to the late Prof.\ Walls for his support, inspiration,
and encouragement during his last years.
This work was
financially supported by the
EC through the TMR Network ERBFMRXCT96-0066.

\end{document}